\documentclass{article}
\usepackage{ijcai22}

\usepackage{times}
\usepackage{soul}
\usepackage{url}
\usepackage[hidelinks]{hyperref}
\usepackage[utf8]{inputenc}
\usepackage[small]{caption}
\usepackage{graphicx}
\usepackage{amsmath}
\usepackage{amsthm}
\usepackage{booktabs}
\usepackage{algorithm}
\usepackage{balance} 
\usepackage{amssymb}
\usepackage{tablefootnote}
\usepackage{framed} 
\usepackage{color}
\usepackage[noend]{algpseudocode}
\usepackage{algorithmicx}
\usepackage{makecell}
\usepackage{multirow} 
\usepackage{subfigure}

{\bfseries}{\itshape}
{\bfseries}{\itshape}
{\bfseries}{\itshape}

\newtheorem{example}{Example}
\newtheorem{theorem}{Theorem}
\newtheorem{lemma}{Lemma}
\newtheorem{definition}{Definition}

\title{Exploring Leximin Principle for Fair Core-Selecting Combinatorial Auctions: Payment Rule Design and Implementation}

\author{
Hao Cheng$^{1}$ \thanks{The work was done during a visiting scholar program at Nanyang Technological University.}\and
Shufeng Kong$^2$ \thanks{Corresponding author}\and
Yanchen Deng$^3$\and
Caihua Liu$^4$\and
Xiaohu Wu$^5$\and\\
Bo An$^3$ \and
Chongjun Wang$^{1}$
\\
\affiliations
$^1$State Key Laboratory for Novel Software Technology, Nanjing University, China\\
$^2$School of Software Engineering, Sun Yat-Sen University, China\\
$^3$School of Computer Science and Engineering, Nanyang Technological University, Singapore \\
$^4$School of Artificial Intelligence, Guilin University of Electronic Technology, China\\
$^5$National Engineering Research Center of Mobile Network Technologies, Beijing University of Posts and Telecommunications, China
\emails
chengh@smail.nju.edu.cn, kongshf@mail.sysu.edu.cn, ycdeng@ntu.edu.sg, caihua.liu@guet.edu.cn, xiaohu.wu@bupt.edu.cn, boan@ntu.edu.sg, chjwang@nju.edu.cn}

\begin{document}

\maketitle

\begin{abstract}

Core-selecting combinatorial auctions (CAs) restrict the auction result in the core such that no coalitions could improve their utilities by engaging in collusion. The minimum-revenue-core (MRC) rule is a widely used core-selecting payment rule to maximize the total utilities of all bidders. However, the MRC rule can suffer from severe unfairness since it ignores individuals' utilities. To address this limitation, we propose to explore the leximin principle to achieve fairness in core-selecting CAs since the leximin principle prefers to maximize the utility of the worst-off; the resulting bidder-leximin-optimal (BLO) payment rule is then theoretically analyzed and an effective algorithm is further provided to compute the BLO outcome. Moreover, we conduct extensive experiments to show that our algorithm returns fairer utility distributions and is faster than existing algorithms of core-selecting payment rules.

\end{abstract}


\maketitle 


\section{Introduction}
Combinatorial auctions (CAs) have found applications in many real-world auction problems, including procurement auctions  \cite{sandholm2007expressive} and government spectrum auctions \cite{cramton2013spectrum,ausubel2017practical,leyton2017economics}. CAs are attractive as they can produce efficient outcomes even when bidders have complex preferences on bundles of heterogeneous goods. A combinatorial auction mechanism often includes two components: 1) an allocation rule to specify the bundle of items assigned to each bidder; 2) a payment rule to specify the price a bidder should pay for his bundle. 
Note that an allocation rule is normally obtained by solving the NP-hard winner determination problem \cite{rothkopf1998computationally}. In this paper, we will focus on the payment rules' designs and implementations. 

To address the truthfulness issue of bidding, where bidders are unwilling to give their real valuations of items but report false information to improve their utilities, the well-known Vickrey-Clarke-Groves (VCG) payment rule \cite{vickrey1961counterspeculation,clarke1971multipart,groves1973incentives} ensures that the optimal strategy for each bidder is to bid their true valuations of the items. However, the VCG payment rule still suffers from various other issues that prohibit it from real-world adoption, e.g., arbitrarily low revenue for the seller, which may cause collusion between the seller and some bidders \cite{ausubel2006lovely}. 

The drawbacks of the VCG payment rule have motivated the development of core-selecting payment rules \cite{day2007fair}, which provide a principled way to ensure that there are no opportunities for collusion between the seller and bidders. The core is a set of all possible payment outcomes where the utility of any coalition of bidders in the auction cannot be improved by engaging in collusion \cite{gillies1959solutions}. 
 However, the core can be exponentially large, and thus, selecting a suitable payment outcome from the core is a critical problem in core-selecting CAs. 
 Many efforts have been devoted to solving this problem by designing some principles of selection \cite{day2007fair,erdil2010new,day2012quadratic,lubin2015new,bunz2018designing}. 
 Among them, the minimum-revenue-core (MRC) rule is particularly popular, where it achieves the largest total incentive by selecting the core outcome that maximizes the total utilities of all bidders.
 
 However, it is observed that the MRC rule may cause severe unfairness, since some winners may be forced to sacrifice their utilities in the MRC rule. A detailed example to illustrate this unfair phenomenon caused by the MRC rule is provided in Example \ref{example} of this paper, where, in the optimal MRC payment outcome, a winner's utility is $0$ while he could have obtained a positive utility. More importantly, this zero-utility phenomenon is not rare since nearly 30\% of winners are forced to get a utility of zero on average in our data generated by CATS \cite{leyton2000towards}, which simulates bidding behavior in many realistic economic environments. 

To address the limitation of the MRC rule, we propose to explore the leximin principle to achieve fairness in core-selecting CAs and propose a novel bidder-leximin-optimal (BLO) payment rule that selects a core outcome based on the leximin principle which prefers to maximize the utility of the worst-off \cite{rawls2004theory}. Informally, the BLO rule maximizes the minimum utility, the second minimum utility and so forth among all the bidders, which achieves egalitarian fairness. Furthermore, we analyze the BLO outcome theoretically and provide an effective algorithm to compute it. We summarize our contributions\footnote{Note that we will refer all proofs to the appendix. Our appendix and code are  available at \url{https://github.com/ha0cheng/Fair-BLO}.} as follows:

\begin{itemize}
    \item[1.] We investigate the unfair phenomenon of the MRC rule and propose the BLO rule to achieve fairness in the core-selecting CAs. 
    \item[2.] We give a theoretical analysis of the BLO rule and show that the BLO outcome is unique and Pareto optimal. Moreover, we offer that the worst-case relative ratio between a bidder's utility (resp. the total utility of all bidders) in the BLO outcome and the maximum utility among all the core outcomes is $\frac{1}{|W|}$ (resp. $\frac{4}{|W|+2+\frac{|W|\mod\;2}{|W|}}$), where $|W|$ is the winner size. 
    \item[3.] Finally, we propose an exact algorithm named {WF-CGS-CR} to compute the BLO outcome. Our {WF-CGS-CR} algorithm consists of the water-filling framework, a constraint generation search subroutine, and a constraint-reuse strategy. Given the oracle access for winner determination, it is effective in time polynomial in the number of winners. Extensive experiments show that our algorithm returns fairer utility distributions and is also faster than existing algorithms of core-selecting payment rules.
    
\end{itemize}

\section{Related Work}
As a practical mechanism, core-selecting CAs and the MRC rule were proposed in \cite{day2007fair} and used in several scenarios, including selling wireless spectra \cite{cramton2013spectrum} and online advertising \cite{goel2015core}. Then the quadratic rule \cite{day2012quadratic} was proposed to select the MRC point nearest to the VCG point in $l_2$-distance. They also study some other reference points like all-zero points. Recently, some works design the core-selecting payment rule by approximating Bayesian equilibrium of different core outcomes \cite{lubin2015new,bunz2018designing}.

Except for the incentives, fairness is an essential metric in the core-selecting mechanisms. \cite{lubin2015new} studied the fairness measured by the Gini coefficient and proposed a fractional rule that selects a point fraction to the VCG point performs best. However, their experiments show that relaxing the MRC constraint does not change the equilibrium properties; and there are no systematic shifts in efficiency, fairness, and aggregate incentives. There are many principles to define fairness. In this paper, we adopt the leximin principle in the core-selecting CAs. The method is similar to the egalitarian core proposed by Arin \cite{arin2001egalitarian}, which selects the utility distribution in the cooperative game via the leximin principle \cite{rawls2004theory}. The difference is that the total utility is not constant in core-selecting CAs.  

As for the pricing algorithms, given access to the separation oracle of winner determination, one can use the Ellipsoid algorithm \cite{grotschel1981ellipsoid} to optimize convex objectives (e.g., bidders' total utility) exactly over the core polytope in polynomial time. However, this approach is rather slow in practice, and Core Constraint Generation (CCG) algorithm \cite{day2007fair} was proposed as an effective heuristic algorithm to solve such an optimization problem, which is commonly used in core-selecting CAs \cite{day2012quadratic,cramton2013spectrum,bunz2015faster}. 

However, it is hard to formulate a global objective for the BLO rule, making it difficult to use the above algorithms directly. Then, Fast Core algorithm \cite{niazadeh2022fast} was proposed to compute an approximate Pareto optimal outcome in the rich advertisement, which can be used to compute an approximate BLO core outcome. Instead of the approximate algorithm, we focus on the exact algorithm. Moreover, the experiments show that our exact algorithm is much faster than Fast Core.

\section{Preliminaries}
\label{Preliminary}

\emph{Combinatorial auctions} are specifically designed for domains where bidders can have complex preferences over bundles of heterogeneous items. Formally, let $N=\{1,\ldots,n\}$ be a set of bidders and $s$ be the seller. Let $M=\{a,b, c, \ldots\}$ be the set of items with $|M|=m$. Each bidder $i$ has a valuation function $v_i:2^M\rightarrow \mathbb{R}$ which specifies the bidder's valuation for every possible bundle of items $S\in 2^M$. Unless stated otherwise, bidders will only need to give their values for bundles in which they are interested, and all other bundles will be given values $0$. Also, we assume that $v_i(\emptyset)=0\;(\forall i \in N)$.  

A CA consists of two steps: step 1 computes an item allocation outcome, and step 2 computes an item payment outcome. An allocation rule specifies the bundle of items $a_i$ assigned for each bidder $i$, where $a_i\in 2^M$ and $a_i \cap a_j = \emptyset \;(\forall i \ne j)$, and a payment rule specifies the price $p_i$ that the bidder $i$ should pay for his bundle $a_i$. Obviously, $p_i$ must be $0$ if $a_i$ is empty.

Let $\pi_i$ denote the utility of an auction participant $i\in N\cup \{s\}$. If $i$ is the bidder, we assume that his utility is quasi-linear, i.e.,  $\pi_i = v_i(a_i)-p_i$. If $i$ is the seller $s$, his utility is  $\pi_s=\sum_{i\in N}p_i$, which is also known as the revenue.  The \emph{social welfare} is the summation of all participants' utility: 
\begin{equation}\label{eq:max-social-welfare}
    \pi_s + \sum_{i\in N}\pi_i =  \sum_{i\in N}p_i +  \sum_{i\in N}(v_i(a_i)-p_i) = \sum_{i\in N} v_i(a_i) 
\end{equation}

One can observe that the social welfare does not depend on the payment, and an allocation rule can be computed by maximizing the social welfare as follows:
\begin{equation}\label{eq:max-utility-seller}
   \max_{\{a_i|a_i\subseteq M, i\in N, a_i\cap a_j =\emptyset, \forall i\neq j\} }\sum_{i\in N}v_i(a_i)  
\end{equation}

Note that the above optimization problem is also known as the \emph{winner determination problem}. Let $ w(\{v_i(\cdot)\}_{i\in N})$ be the maximum social welfare given the valuation function profile $\{v_i(\cdot)\}_{i\in N}$.
Once we have a social welfare maximizing allocation $\{a_i|i\in N, a_i\cap a_j =\emptyset, \forall i\neq j\}$ by solving the winner determination problem, in step 2, we need to decide the price $p_i$ for each allocated item bundle $a_i(i\in N)$. 

The simplest payment rule is the so-called \emph{pay-as-bid} payment rule, where a winner $i (a_i\ne \emptyset)$ pays the price $p_i=v_i(a_i)$ for buying the item bundle $a_i$. However, the pay-as-bid payment rule suffers from the truthfulness issue, where bidders are unwilling to report the true valuation functions but strategically give false information instead. Therefore, in step 2, the famous VCG mechanism is proposed to offer bidders incentives to ensure that bidding truthfully is the dominant strategy. Unfortunately, VCG can lead to very low or even zero revenue, which is unstable and may cause collusion between the seller and some bidders.

\subsection{Core-selecting Payment Rules}

The weaknesses of VCG payment rule have motivated the development of \emph{core-selecting payment rules},  which provide a principled way to ensure that the revenue of the auction is high enough such that there are no opportunities for collusion. 

Formally, let $S \subseteq N\cup \{s\}$ denote a coalition of participants.Then, if $s\not\in S$, the social welfare of $S$ must be $0$ since the auction cannot happen without the seller's participation; If $s\in S$, the social welfare of $S$ can be computed as follows: 
\begin{equation}
\small
   \pi_s+\sum_{i\in S\setminus \{s\}}\pi_i = \sum_{i\in S\setminus \{s\}}p_i + \sum_{i\in S\setminus \{s\}}(v_i(a_i)-p_i)  = \sum_{i\in S\setminus \{s\}}v_i(a_i) 
\end{equation}

The coalition characteristic function, $\mathcal{F}:2^{N\cup\{s\}}\rightarrow \mathbb{R}$, which indicates the optimal social welfare of each coalition $S$ is then defined as follows:
\begin{equation}
    \mathcal{F}(S) = \left\{
    \begin{aligned}
    &    w(\{v_i(\cdot)\}_{i\in S\setminus \{s\}}), \text{if}\; s\in S  \\
    &0, \text{if}\; s\notin S \\
    \end{aligned}
    \right.
\end{equation}

The \emph{core} is the set of all possible outcomes where the total utility of any coalition cannot be improved by engaging in collusion \cite{gillies1959solutions}. The definition is given as follows:
\begin{definition}[Core]
 The core of a CA is a set of outcomes satisfying the following two conditions:
 
\begin{itemize}
    \item Efficiency:
\begin{equation}\label{eq:Efficiency}
     \sum_{i\in N\cup \{s\}} \pi_i = \mathcal{F}(N\cup \{s\})
\end{equation}
    \item Coalitional rationality:
    \begin{equation}\label{eq:Coalitional-rationality}
     \sum_{i\in S} \pi_i \geq  \mathcal{F}(S) ,\;\forall S\subseteq N\cup\{s\}
\end{equation} 
\end{itemize}
\end{definition}

Note that the efficiency condition requires that the  allocation outcome must be optimal, namely, the allocations are obtained by solving the winner determination problem. It is possible that there is more than one optimal allocation, and we break ties by selecting a random optimal allocation $\{a^*_i\}_{i\in N}$. Denote by $W$ the winner set, i.e., $W=\{i|a^*_i\ne \emptyset\}$. On the other hand, the coalitional rationality condition guarantees that the maximum social welfare that any coalition $S$ could get (i.e., RHS of the inequality) is not larger than what the coalition can already get under the current outcome (i.e., LHS of the inequality). 
\emph{Finally, a core-selecting CA selects a social welfare maximizing allocation and picks a payment outcome in the core.}

If coalition $S$ does not include the seller, then $\mathcal{F}(S)=0$. Thus the corresponding constraints of coalitional rationality can be summarized as follows:
\begin{equation}\label{eq:Indivual Rational}
    \pi_i\geq 0,\;\forall i\in N
\end{equation}

 If $s\in S$, substitute Eq.\ref{eq:Efficiency} into Eqs.\ref{eq:Coalitional-rationality} and we can arrange the constraints as follows:
 \begin{equation}\label{eq:constraints}
     \sum_{i\in N\setminus S}\pi_i\leq w(N) - w(S),\;\forall S\subseteq N
 \end{equation}
where $w(S)$ is the abbreviation of $w(\{v_i(\cdot)\}_{i\in S})$. Eqs.\ref{eq:Indivual Rational} and  Eqs.\ref{eq:constraints} form the complete core constraint set, which includes $2^N+N$ constraints in total.
We use the utility to analyze in the following discussion and each bidder's payment price is mapped by $p_i=v_i(a^*_i)-\pi_i$.  

\subsection{Minimum-Revenue-Core Payment Rule}
\emph{Core-selecting CAs} are not guaranteed to be truthful \cite{goeree2016impossibility}. To maximize the incentives of truthful bidding, the \emph{Minimum-Revenue-Core rule} (MRC) is proposed as a fundamental principle to select the core outcomes \cite{day2007fair,day2008core,day2012quadratic}. The MRC rule maximizes the total utility of all bidders, which can be computed as follows:
\begin{equation}
    \max_{\{\pi_i\}_{i\in N}\in U} \sum_{i\in N}\pi_i
\end{equation}\
where $U$ is the core of the related CA. 

Unfortunately, we observe that the MRC rule could suffer from severe unfairness of utility distribution, where some winners' utilities are forced to be zero. This unfair phenomenon largely dues to the fact that the MRC rule emphasizes the total utility and ignores individuals' utilities.  We will illustrate this unfair utility distribution phenomenon in the following example. 

\begin{example}\label{example}
Consider a CA with $5$ bidders and $3$ items. Let $N=\{1,2,3,4,5\}$ be the set of bidders and $M=\{a,b,c\}$ be the set of items. Table \ref{Toy example} gives the valuation functions of all bidders. By solving the winner determination problem, we have that the winners are bidders 1, 2, and 3, thus the maximum social welfare is that $w(N) = v_1(\{a\})+v_2(\{b\})+v_3(\{c\})=6$ according to Eq.\ref{eq:max-social-welfare}. Similarly,  $w(\{3,4,5\})=4$ and the corresponding constraint is $\pi_1+\pi_2\leq 6-4=2$. The other constraints are obtained similarly. After removing the redundant constraints, the core of this example is shown as follows:
\begin{equation}\label{eq:example core}
      \text{Core} = \left\{
    \begin{aligned}
    &\pi_1 \geq 0, \pi_2 \geq 0,\pi_3 \geq 0\\
    &\pi_1 + \pi_2 \leq 2 \\
    &\pi_2 + \pi_3 \leq 2\\
    \end{aligned}
    \right.  
\end{equation}

According to the MRC rule, the unique utility distribution is $\{2,0,2\}$. Bidder 2 must sacrifice all of his utility even though he could have obtained a positive utility in other core outcomes (e.g., $\{1,1,1\}$). In this case, bidder 2 would prefer bidding untruthfully to get more utility. For example, he can bid the value $1$ for the item $b$, which leads to a utility of $1$ in the MRC rule.  
\end{example}

\begin{table}[htbp]
\caption{A CA with 5 bidders and 3 items. Winning bids are marked with ``*''. The last three columns are utilities computed from the VCG, MRC, and our new BLO rule, respectively. }
\setlength\tabcolsep{3pt}
    \centering
    \small
    \begin{tabular}{l|ccccc|ccc} 
    \hline
    &\multicolumn{5}{c|}{bids}&\multicolumn{3}{c}{utilities} \\
    & \{a\} & \{b\} & \{c\}& \{a,b\} & \{b,c\} & VCG & MRC & BLO  \\
        \hline
        Bidder 1 & 2*  &     &   &   &  & 2&  2 & 1\\
        Bidder 2 &   &2* &&&& 2& 0  & 1\\
        Bidder 3 &  &&2* && & 2&  2 & 1\\
        Bidder 4 & &&&2&&  0&0   &0 \\
        Bidder 5 & &&&&2 &  0&0 &0 \\ 
    \hline
    \end{tabular}
    \label{Toy example}
\end{table}

Extending this example to the general case, we have the following lemma:
\begin{lemma}\label{lem:unfairness}
     Assuming that bidders bid truthfully, in the worst case, $|W|-2$ winners may get zero utility in the MRC rule, even though there exists some core outcome that assigns them with positive utilities, where $|W|$ is the winner size.
\end{lemma}

Moreover, we observe that the unfair phenomenon often happens in a wide range of domains and datasets when using the MRC rule. To address this issue, we will explore and propose a novel payment rule to achieve fairness. 


\section{Bidder-Leximin-Optimal Payment Rule}

We propose to adopt the \emph{leximin principle} to achieve fairness in core-selecting CAs, where the leximin principle prefers the utility distribution with a larger utility for the worst-off. 

\begin{definition}[Leximin Dominance]
    Let $\boldsymbol{x},\boldsymbol{y}\in \mathbb{R}^n$ be two real vectors. $\boldsymbol{x}$  leximin dominates $\boldsymbol{y}$ if the following holds:
    \begin{itemize}
        \item There exists some integer $0\le k\le n-1$ such that the $k$-smallest elements of both vectors are equal, and the $(k+1)$-smallest element of $\boldsymbol{x}$ is larger than the $(k+1)$-smallest element of $\boldsymbol{y}$.
    \end{itemize}
\end{definition}

We denote the leximin-dominant relation by ``$\succ_{Lex}$'', and then a \emph{Bidder-Leximin-Optimal} (BLO) core outcome is defined as follows:

\begin{definition}[Bidder-Leximin-Optimal]
Given a core $U$ of a CA, a utility outcome $\boldsymbol{\pi} \in U$ is bidder-leximin-optimal if and only if it is not leximin dominated by any utility outcome $\boldsymbol{\pi}'$ in the core, i.e., 
\begin{equation}
\not\exists \boldsymbol{\pi}' \in U: \boldsymbol{\pi}' \succ_{Lex} \boldsymbol{\pi}
\end{equation}
\end{definition}

The BLO payment rule follows the leximin principle to give priority to those who are worst-off. It selects a core outcome that is BLO rather than a core outcome that maximizes the total utility of bidders. In order to illustrate the advantage of the BLO payment rule over the MRC payment rule, we consider the following example.

\begin{example}
    Let us recall the CA problem in example 1. The unique MRC utility outcome is $\pi=\{2,0,2\}$ for the winners. On the other hand, the BLO utility outcome is $\boldsymbol{\pi}'=\{1,1,1\}$ under the core constraints in Eqs.\ref{eq:example core}. One can observe that $\boldsymbol{\pi}'$ is a fairer utility distribution than $\boldsymbol{\pi}$. 
\end{example}

\subsection{Theoretical Analysis of the BLO Payment Rule}
This section provides three theoretical results for the BLO payment rule. First, the following theorem establishes that the BLO outcome is unique and Pareto optimal.

\begin{theorem}\label{thm:unique of BLO}
The BLO outcome always exists, and it is unique and Pareto optimal, where no bidders can improve their utilities without harming other bidders' utilities.
\end{theorem} 

We then consider a lower bound of a bidder's utility in the BLO core outcome.
\begin{theorem}\label{thm:single utility in BLO}
Assuming that bidders bid truthfully, given the BLO outcome $\{\pi^{BLO}_i\}_{i\in N}$, we have the following tight lower bound:
\begin{equation*}
\pi^{BLO}_i \geq \frac{1}{|W|}\pi^*_i
\end{equation*}
where $\pi^*_i$ is the maximum utility for bidder $i$ in the core. 
\end{theorem}

The above lower bound shows that a bidder's utility is always larger than zero in the BLO outcome, unless it is zero in every core outcome's utility distribution. Regarding the total utility of the BLO outcome, it is clear that it cannot be better than that of an MRC outcome, because the total utility of an MRC outcome is maximum. The following result shows the ratio between the BLO outcome and the MRC outcome in terms of their total utility.

\begin{theorem}\label{thm:total utility in BLO}
Assuming that bidders bid truthfully, given the BLO utility outcome $\{\pi^{BLO}_i\}_{i\in N}$, we have the following tight lower bound:
\begin{equation*}
 \sum_{i\in N} \pi^{BLO}_i\geq \frac{4}{|W|+2+\frac{|W|\mod\;2}{|W|}} \sum_{i\in N} \pi^{MRC}_i
\end{equation*}
where $\{\pi^{MRC}_i\}_{i\in N}$ is an MRC outcome.
\end{theorem} 

One can observe that the relative ratio is 1 if the size of the winner set is 1 or 2, but the relative gap grows linearly with the winner set size $|W|$. Besides, the BLO outcome has a better guarantee than the general Pareto optimal outcome, which has a tight relative ratio guarantee of $\frac{1}{|W|-1}$ \footnote{More detail in the appendix.}.

\section{An Exact Algorithm for the BLO Payment Rule}
\label{WF}

In this section, we propose an exact algorithm, named {WF-CGS-CR}, to compute the BLO payment outcome. The {WF-CGS-CR} algorithm is a \textbf{water-filling (WF) algorithm} to increase all active winners' utilities uniformly in an interactive manner until reaching the BLO outcome, and a \textbf{constraint generation search (CGS) subroutine} is adopted to compute the maximum utility increment at each water-filling iteration. Moreover, we offer a \textbf{constraint-reuse (CR) strategy} to accelerate the CGS by setting a better initialization with the history of constraints. We will elaborate on each component of our {WF-CGS-CR} below.

Since core-selecting CAs involve solving the winner determination problem, existing works \cite{day2007fair,day2012quadratic} often assume that there is a so-called winner determination oracle to solve this problem automatically. We will follow such an assumption in this paper and introduce the concept as follows:

\begin{definition}[$\mathcal{WD}$]
Let $\mathcal{WD}$ be an oracle with this input-output relation:\\
\noindent\textbf{Input}: submitted bid profile $\{b_i(\cdot)\}_{i\in N}$, where $b_i:2^M\rightarrow \mathbb{R}$ is bidder $i$'s reported valuation function\footnote{We regard the bidders' reported valuation functions as their true valuation functions in the computation part.}.\\
\noindent\textbf{Output}: a winner set $W$ and maximum social welfare $w(\{b_i(\cdot)\}_{i\in N})$ under the submitted bids.
\end{definition}

In other words, we do not dig into the winner determination solving but only regard it as a black-box interface. Any state-of-the-art winner determination method can be used in this way. \textit{Oracle complexity} is defined as the query time to the oracle in the worst case, which is the primary analysis aspect for the algorithms in this paper.

\subsection{Water-filling Algorithm}
Water-filling algorithm was first proposed in \cite{niazadeh2022fast} to compute an approximate Pareto optimal outcome. Different from the original version, we develop the algorithm for the exact BLO outcome.

Given a utility distribution in the core, we have two kinds of winners: the active winner and the frozen winner. The active winner has the potential to increase his utility, while the frozen winner can not increase the utility further. Our water-filling algorithm starts from the all-zero utility distribution and the initial active winner set $W$ (i.e., the original winner set). Then, at the $t$-th iteration, it does the following: 
\begin{itemize}
    \item Run the search subroutine to return the maximum utility increment $\Delta\pi^t$ and the frozen winner set $W^{t}_f$. 
    \item Generate the next utility distribution by increasing each active winner's utility by  $\Delta\pi^t$ and generate the next active winner set by removing $W^t_f$ from the current active winner set $W^t_a$.
\end{itemize}

The iteration stops when the active winner set becomes empty, i.e., no one can increase his utility anymore. The pseudo-code is shown in Alg.\ref{alg:WF}. 

\begin{algorithm}[htbp]
  \caption{Water-filling algorithm for finding the exact BLO outcome}\label{alg:WF}
    \textbf{Input}:  bid profile $\{b_i(\cdot)\}_{i \in N}$.    \\
      \textbf{Output}: the BLO outcome.
    \begin{algorithmic}[1]
     \State Step $t\leftarrow 1$
     \State $W,w(N) \leftarrow \mathcal{WD}(\{b_i(\cdot)\}_{i\in N})$
    \State Initialize the utility distribution $\{\pi^t_i\}_{i\in N}$ as all-zero 
    \State Initialize the active winner set $W^t_a$ as $W$
      \While{$W^t_a$ is not empty}
      \State Run the search subroutine to return the maximum utility increment $\Delta\pi^t$ and the frozen winner set $W^{t}_f$
        \State Generate the next utility distribution:
        \begin{equation*}
        \pi^{t+1}_i = \left\{
        \begin{aligned}
        &\pi^t_i +\Delta\pi^t,\; \text{if $i\in W^t_a$}\\
        & \pi^t_i,\; \text{otherwise}
        \end{aligned}
        \right.
    \end{equation*}
        \State Generate the next active winner set: 
        \begin{equation*}
        W^{t+1}_a = W^t_a \backslash W^{t}_f
        \end{equation*}
        \State Step $t\leftarrow t+1$
      \EndWhile
      \State \Return{$\{\pi^t_i\}_{i\in N}$}
  \end{algorithmic}
\end{algorithm} 

A search subroutine is adopted to find the maximum utility increment and the frozen winner set at each iteration. Formally, given that the current utility distribution $\{\pi^t_i\}_{i\in N}$ is in the core, the optimization problem for the search subroutine can be formulated as follows:
 \begin{equation}\label{P0}\tag{P0}
\max_{\{\Delta\pi| \{\pi^t_i+\Delta\pi\}_{i\in W^t_a}\in U\}}\Delta\pi
\end{equation}
where $\{\pi^t_i+\Delta\pi\}_{i\in W^t_a}$ means increasing the utilities of winners in $W^t_a$ by $\Delta\pi$ while the other winners' utilities remain unchanged. After the utility increment, some winners' utilities can not increase anymore, i.e., the frozen winner set at this iteration. Then we have the following lemma:
\begin{lemma}\label{lem:water-filling}
Given that the results computed by the search subroutine are correct, the water-filling algorithm returns the BLO outcome.
\end{lemma}

\subsection{Constraint Generation Search Subroutine}
\label{section:PF}
A naive method to solve the problem (\ref{P0}) is obtaining all the core constraints, but it needs to query the winner determination oracle exponential times according to Eqs.\ref{eq:constraints}. To avoid this, in this section, we first formulate the problem (\ref{P0}) to an equivalent mathematical format and then solve it through the constraint generation search.

\subsubsection{Optimization Problem Formulation}

In the original format (\ref{P0}), the restriction condition is to ensure the utility distribution is still in the core, in other words, satisfying all the core constraints. According to Eqs.\ref{eq:constraints}, the constraints can be written as:
\begin{equation}
    \sum_{i \in N\setminus S}\pi^t_i + |W^t_a\setminus S|\Delta\pi \leq w(N) -  w(S),\;\forall S\subseteq N
\end{equation}
    
 If $W^t_a\subseteq S  $, then $|W^t_a\setminus S|=0$, and the corresponding constraints can be ignored. Thus, we can arrange the remaining constraints as follows:
\begin{equation} \label{eq:upper bound}
    \Delta\pi \leq \frac{w(N)  - w(S) - \sum_{i \in N\setminus S}\pi^t_i}{|W^t_a \backslash S|}
\end{equation}
where $S\subseteq N \;and \;W^t_a \not\subseteq S$. Every such constraint provides an upper bound for $\Delta\pi$.  Denote by $\Delta\pi^S$ the upper bound corresponding to $S$, i.e., RHS in  Eq.\ref{eq:upper bound}. Thus, finding the maximum utility increment satisfying all the core constraints is equivalent to finding the minimum upper bound above. Problem (\ref{P0}) is equivalent to the following optimization problem:
\begin{equation}\label{eq:Problem}\tag{P1}
     \min_{\{S|S\subseteq N,W^t_a \not\subseteq S\}}\frac{w(N)  - w(S) - \sum_{i \in N\setminus S}\pi^t_i}{|W^t_a \backslash S|}
\end{equation}

Let $S^*$ be one of the optimal coalitions for problem (\ref{eq:Problem}), thus $\Delta\pi^t=\Delta\pi^{S^*}$. After the increment, the corresponding core constraint of $S^*$ would be tight for the new utility distribution $\{\pi^{t+1}_i\}_{i\in N}$, thus we have
\begin{equation}\label{eq:1}
    \sum_{i \in N\setminus S^*}\pi^{t+1}_i  = w(N) - w(S^*)
\end{equation}
According to Eq.\ref{eq:1}, bidders belonging to $N\backslash S^{*}$ cannot increase their utilities without violating this constraint, i.e., they are frozen. Therefore, the frozen winner set $W^t_f$ is $W^t_a \backslash S^{*}$ at this iteration. 
Problem (\ref{eq:Problem}) is actually a nesting optimization problem with a fractional format since $w(S)$ is corresponding to the winner determination problem. Thus it is intractable to solve directly, and we propose to use the constraint generation method that only considers the most useful constraints.

\subsubsection{Constraint Generation Search}
Generally speaking, CGS starts from an upper bound of the utility increment, then reduces the upper bound iteratively through generated constraints until it is the minimum.

Formally, CGS initializes the upper bound $\Delta\bar{\pi}$ as $\infty$ and the frozen winner set as $W^t_a$. It checks whether this upper bound is the minimum by the following equation:
\begin{equation}\label{eq:in core condition}
    w(N) - \sum_{i\in N}\tilde{\pi}_i = w_B
\end{equation}
 where $\{\tilde{\pi}_i\}_{i\in N}$ is the utility distribution by setting the utility increment as $\Delta\bar{\pi}$ and $w_B$ is the maximum social welfare under the truncated bid profile $\{\max(b_i(\cdot)-\tilde{\pi}_i,0)\}_{i\in N}$. If Eq.\ref{eq:in core condition} is satisfied, the current upper bound $\Delta\bar{\pi}$ is the minimum, and CGS will return the result. Otherwise, a smaller upper bound is computed by
    \begin{equation}\label{eq:upper_bound}
       \Delta\bar{\pi} =  \Delta\bar{\pi} - \frac{w_B-(w(N) - \sum_{i\in N}\tilde{\pi}_i)}{|\bar{W}_f|}
    \end{equation} 

The corresponding frozen winner set is updated by 
\begin{equation}\label{eq:frozen winner set}
        \bar{W}_f=W^t_a\backslash B
\end{equation}
where $B$ is the winner set under the truncated bid profile $\{\max(b_i(\cdot)-\tilde{\pi}_i,0)\}_{i\in N}$. If the new frozen winner set satisfies $|\bar{W}_f|=1$, then this upper bound is the minimum. Otherwise, CGS will start a new iteration with the smaller upper bound. The oracle complexity of CGS is given as follows:

\begin{lemma}\label{lem:CGS}
 Constraint generation search requires at most $|W^t_a|$ queries to the oracle $\mathcal{WD}$ to solve the optimization problem (\ref{eq:Problem}). 
\end{lemma}

Assume that the iteration number of the water-filling algorithm is $T$. Naturally, since $W =  W^1_a\supset...\supset W^T_a=\emptyset$, the query time of the CGS subroutine in the water-filling algorithm is at most $|W|+(|W|-1)+...+1=\frac{|W|(|W|+1)}{2}$.

 \begin{algorithm}[!t]
  \caption{Constraint generation search with the constraint-reuse strategy}\label{alg:CGS}   
    \textbf{Input:}  
      bid profile $\{b_i(\cdot)\}_{i \in N}$, maximum social welfare $w(N)$, current utility distribution $\{\pi^t_i\}_{i\in N}$, active winner set $W^t_a$, last utility increment $\Delta\pi^{t-1}$, constraint set $\mathbf{C}$. \\
      \textbf{Output:}  the maximum utility increment $\Delta\pi^t$ and the frozen winner set $W^t_f$. 
  \begin{algorithmic}[1]  
        \Statex /*Constraint-reuse Strategy*/
        \State Initialize $\Delta\bar{\pi}\leftarrow \infty;\;\bar{W}_f\leftarrow W^t_a$
        \For{$(W^S_f,\Delta\pi^S)$ in $\mathbf{C}$}
        \State $\mathbf{C} \leftarrow \mathbf{C}\setminus \{(W^S_f,\Delta\pi^S)\}$
        \If{$W^S_f\cap W^t_a \neq \emptyset$} 
        \State$\acute{W}^S_f \leftarrow W^S_f\cap W^t_a;\;\Delta\acute{\pi}^S = \frac{|W^S_f|}{|\acute{W}^S_f|}(\Delta\pi^S - \Delta\pi^{t-1})$
        \If{$\Delta\acute{\pi}^S<\Delta\bar{\pi}$}
        $\Delta\bar{\pi}\leftarrow \Delta\acute{\pi}^S$;
        $\bar{W}_f\leftarrow \acute{W}^S_f$ 
        \EndIf
        \State $\mathbf{C} \leftarrow \mathbf{C}\cup\{(\acute{W}^S_f,\Delta\acute{\pi}^S)\}$
        \EndIf
        \EndFor
        \Statex /*Constraint Generation Search*/
        \While{True}
        \State Generate the utility distribution by upper bound $\Delta\Bar{\pi}$:      
        \begin{equation*}
        \tilde{\pi}_i = \left\{
        \begin{aligned}
        &\pi^t_i +\Delta\bar{\pi},\; \text{if $i\in W^t_a$}\\
        & \pi^t_i,\; \text{otherwise}
        \end{aligned}
        \right.
        \end{equation*}
        \State  $B,w_B \leftarrow \mathcal{WD}(\{\max(b_i(\cdot)-\tilde{\pi}_i,0)\}_{i\in N})$ 
        \If{ $w(N) - \sum_{i\in N}\tilde{\pi}_i = w_B$ } break
        \EndIf
        \State $\Delta\bar{\pi} \leftarrow \Delta\bar{\pi} - \frac{w_B-(w(N) - \sum_{i\in N}\tilde{\pi}_i)}{|\bar{W}_f|};\;\bar{W}_f\leftarrow W^t_a\backslash B$ 
        \State $\mathbf{C} \leftarrow \mathbf{C}  \cup\{(\bar{W}_f,\Delta\bar{\pi})\}$ 
        \If{$|\bar{W}_f|=1$} break
        \EndIf
        \EndWhile
    \State \Return{$\Delta\bar{\pi}$,$\bar{W}_f$} 
  \end{algorithmic}
\end{algorithm}

\subsection{Constraint-reuse Strategy}
\label{section:CR}
To avoid CGS starting from the naive upper bound $\infty$, we propose the constraint-reuse strategy to set a better initialization through the history of constraints.

Formally, we store the binary tuple $(W^S_f,\Delta\pi^S)$ in the stored constraint set $\mathbf{C}$, where $W^S_f$ is the frozen winner set if $S$ is an optimal coalition, i.e., $W^S_f = W^t_a\setminus S$. Then in each CGS subroutine, we first update each binary tuple $\mathbf{C}$ based on the new utility distribution and active winner set. The recurrence formula is given as follows:
\begin{equation}
     \left\{
    \begin{aligned}
     \acute{W}^{S}_f &= W^S_f \cap W^{t}_a   \\
    \Delta\acute{\pi}^S &= \frac{|W^{S}_f|}{|\acute{W}^{S}_f|}(\Delta\pi^S - \Delta\pi^{t-1})
    \end{aligned}
    \right.
\end{equation}  

\setcounter{table}{2}
 \begin{table*}[!t]
\centering
\caption{Results for the run time. Average run times (in seconds) over 50 CA instances are provided, with standard error in the parentheses. Each CA instance is generated by CATS, with 64 goods and 1000 bids. The lowest average run time in each column is marked in \textbf{bold}.}
\label{runtime}
\begin{tabular}{lllllll}
    \hline  
  Algorithm & Arbitrary& Decay(L4) &Matching &Paths&Regions&Scheduling  \\
   \hline
VCG&12.49 \;(4.52)&1.83 (0.87) &\textbf{0.75} (0.03) &\textbf{2.50} (0.18) &2.35 (0.45) &\textbf{0.83} (0.36)\\
MRC&22.09 (10.30)&4.84 (2.35)&2.12 (0.62)&5.70 (1.54)&4.42 (1.40)&2.37 (1.93)\\
MRC-VCG&23.32 (10.86)&4.84 (2.21)&2.23 (0.63)&5.86 (1.51)&4.76 (1.48)&2.90 (2.49)\\
MRC-Zero&23.48 (10.91)&5.17 (2.74)&2.23 (0.64)&5.90 (1.55)&4.87 (1.47)&2.99 (2.79)\\
Fast Core&33.26 (15.24)&6.80 (3.19)&7.67 (1.00)&21.46 (3.07)&7.23 (2.54)&11.48 (8.9)\\
WF-CGS&9.67 (4.85)&3.57 (1.83)&2.98 (0.42)&10.37 (1.56)&2.37 (1.13)&2.36 (0.97)\\
WF-CGS-CR&\textbf{7.15} (3.64)&\textbf{1.81} (0.93)&1.22 (0.18)&3.25 (0.36)&\textbf{1.55} (0.54)&1.58 (1.17)\\
  \hline
\end{tabular}
\end{table*}
Above all, we have introduced the three components of the WF-CGS-CR algorithm and the runtime guarantee is shown as follows:
\begin{theorem}\label{thm:WF-CGS-CR}
The WF-CGS-CR algorithm requires at most $\frac{|W|(|W|+1)}{2}+1$ queries to oracle $\mathcal{WD}$ and an additional time complexity $O(|W|^5)$ to obtain the BLO outcome, where $|W|$ is the winner size.
\end{theorem}

By adopting this formula, we transfer each binary tuple $(W^S_f,\Delta\pi^S)$ to a new binary tuple $(\acute{W}^{S}_f,\Delta\acute{\pi}^S)$ before CGS. Eventually, the minimum upper bound in the constraint set is computed and serves as the initial upper bound for CGS. Besides, each new binary tuple is stored in the constraint set if $\acute{W}^{S}_f$ is not empty. The complete pseudo-code for the CGS subroutine is shown in Alg.\ref{alg:CGS}. Note that this strategy would increase some extra computation complexity that is minor since the winner determination oracle is the main time-consuming part.



\section{Experiments}
\label{Experiments}
In the experiments, we use the CATS (Combinatorial Auction Test Suite) \cite{leyton2000towards} as the CA instance generator. Same as \cite{bunz2015faster}, we choose six representative CATS distributions: Arbitrary, Decay(L4), Matching, Paths, Regions, and Scheduling. For each distribution, we generate 50 CA instances with 64 goods and 1000 bids. All the experiments were run on a high-performance computer with a 3.10GHz Intel core and 16GB of RAM. Besides, the winner determination problem is transformed into a format of integer programming, solved by the solver CPLEX 20.1.0 \cite{manual1987ibm}. The following payment rules are implemented for the comparison:
\begin{itemize}
    \item \emph{VCG}: the VCG payment rule, where each winner's utility is $\pi^{VCG}_i=w(N)-w(N\setminus \{i\})$.
    \item \emph{MRC}: the MRC payment rule that selects one core outcome with the maximum total utility of bidders \cite{day2007fair}.
    \item \emph{MRC-VCG}: a quadratic payment rule that selects the MRC point nearest to the VCG point in $l_2$-distance \cite{day2012quadratic}.
    \item \emph{MRC-Zero}: a quadratic payment rule that selects the MRC point nearest to the all-zero point in $l_2$-distance \cite{day2012quadratic}.
    \item \emph{Fast Core}: the approximate BLO payment rule implemented through the water-filling algorithm with the binary search \cite{niazadeh2022fast}. 
\end{itemize}

We use the state-of-the-art Core Constraint Generation (CCG) algorithm to compute all the MRC outcomes above \cite{day2007fair}. Unfortunately, it has no theoretical guarantee for the query time; in other words, its oracle complexity may be exponential. Instead, the oracle complexity of Fast Core is $O(|W|\log\frac{|W|}{\epsilon})$, where $\epsilon$ is the parameter representing the approximate gap. Same as \cite{niazadeh2022fast}, $\epsilon$ is set as 0.01 in our experiments. 


\subsection{Experimental Results}
The auction results are shown in Table \ref{mechanism}. We can see that core-selecting payment rules enhance the seller's revenue compared to the VCG payment rule. Among the core-selecting payment rules, the BLO rule achieves a fairer utility distribution with a tiny reduction of the total utility (i.e., from 173.65 to 147.17). Moreover, the MRC rules produce an unfair utility distribution, with nearly 30\% winners obtaining zero utility, which impairs the incentives for bidders. Instead, the BLO rule has only 1.69\% zero-utility winners on average, thus providing a guarantee for the worst-off. Fast Core achieves approximate fairness, but there are still 12.71\% zero-utility winners. 
\setcounter{table}{1}
\begin{table}[htbp]
\setlength\tabcolsep{1pt}
\centering
\caption{Results for the auction. The measurements include the revenue, the total utility, the minimum utility (Min), the standard deviation (Std), and the percentage of the zero-utility winners among all winners (Zero ratio). All results are averages over 300 CA instances, with 50 instances for each distribution. Note that $\uparrow$ indicates larger values are better, and $\downarrow$ indicates smaller values are better in terms of fairness. }
\label{mechanism}
\begin{tabular}{lccccc}
    \hline
  Payment rule &Revenue &Total utility&Min $\uparrow$&Std$\downarrow$&Zero ratio$\downarrow$ \\
   \hline
VCG &11661.54&513.04&5.03&13.63&1.69\%\\
MRC-VCG&12000.93&173.65&0.01&10.36&31.63\%\\
MRC-Zero&12000.93&173.65&0.05&9.54&28.57\% \\
Fast Core&12046.95&127.63&0.83&5.94&12.71\%\\
BLO &12027.41&147.17&1.18&6.77&1.69\% \\
  \hline
\end{tabular}
\end{table}

The average run time results are shown in Table \ref{runtime}, where WF-CGS represents our algorithm without the constraint-reuse strategy. Note that we include the  run time for computing VCG prices for the MRC rules since it serves as the initial point in the CCG algorithm \cite{day2007fair}. As we can see, the WF-CGS-CR algorithm performs the best among all the core-selecting algorithms, which is even close to the VCG algorithm. Moreover, the constraint-reuse strategy is effective in our experiments, especially in the distribution of Paths.

\section{Conclusion}
\label{Conclusion}
In this paper, we propose the BLO payment rule to address the unfair issue with the MRC rule. The BLO payment rule adopts the leximin principle to select the core outcome, prioritizing the worst-off bidders. Furthermore, we theoretically analyze the BLO rule and propose the WF-CGS-CR algorithm to compute the exact BLO outcome. Our algorithm has the quadratic oracle complexity and achieves the best run time in the experiments, even close to VCG.

Compared with the MRC rule, the BLO rule has the following advantages: (1) the BLO outcome is unique, avoiding the problem of core outcome selection; (2) the BLO outcome achieves leximin fairness and thus avoids the unfair phenomenon caused by the MRC rule; (3) given the winner determination oracle access, the BLO outcome is computed effectively in time polynomial in the winner size. Therefore, the BLO rule could serve as an alternative payment rule for the core-selecting CAs.

\section{Acknowledgement}
 This research was supported by the National Natural Science Foundation of China (Grant No. 62192783, U1811462), the Collaborative Innovation Center of Novel Software Technology and Industrialization at Nanjing University. The work of Shufeng Kong was partially supported by the “Hundred Talents Program” of Sun Yat-sen University. The work of Caihua Liu was partially supported and funded by the Humanities and Social Sciences Youth Foundation, Ministry of Education of the People's Republic of China (Grant No.21YJC870009). Bo An is supported by the National Research Foundation, Singapore under its Industry Alignment Fund – Pre-positioning (IAF-PP) Funding Initiative. Any opinions, findings and conclusions or recommendations expressed in this material are those of the author(s) and do not reflect the views of National Research Foundation, Singapore.
  



\bibliographystyle{named}
\bibliography{paper.bib}


\end{document}